\documentstyle[12pt]{article}
\begin{document}
\title{The Hall effect in a nonlinear strongly correlated regime}
\author{Artur Sowa\\
109 Snowcrest Trail,
Durham, NC 27707 \\
www.mesoscopia.com }
\date{}
\maketitle
\begin{abstract}

I examine a model for the Hall effect in the strongly correlated regime. It emerges from an approach proposed in my previous articles [e.g. J. Phys. Chem. Solids, 65 (2004), 1507-1515; J. Geom. Phys., in press, cf. math-ph/0409023].  The model entails the magnetic Schr\"{o}dinger equation with a built-in strongly nonlinear feedback loop. Within the resulting picture, we observe breaking down of the symmetries displayed by the familiar linear problem. In particular, the model predicts Hall potential with either the regular or the anomalous bias and, simultaneously, formation of stable charge-stripes. On the technical side, a certain third-order nonlinear ordinary differential equation becomes the cornerstone of analysis. In this letter the focus is on the qualitative aspects of the model.

\end{abstract}

\section{Synopsis}

Analysis presented in this article has the following two-stage structure. 
First, in Section \ref{secHamil}, a certain nonlinear system of equations, involving a Maxwell-type, and a Schr\"odinger-type part, is solved in order to find the electronic Hamiltonian. These equations balance the interaction of the ambient magnetic field with an ensemble of correlated electrons. The type of interaction considered here is uniquely characteristic of the Mesoscopic Mechanics/ Nonlinear Maxwell Theory approach, cf. \cite{sowa1}, \cite{sowa3}, \cite{sowa4}, \cite{sowa5}, and \cite{sowa6}. Once the effective Hamiltonian is determined, it becomes a constituent of the two-dimensional single-particle-type Schr\"odinger equation, which determines electronic properties of the system. This latter stage of analysis is carried out in Section \ref{predic}. 

\section{The effective Hamiltonian}  

\label{secHamil}
Consider a planar system whose electronic properties are captured with the single-particle Schr\"odinger equation
 \begin{equation}
 i\hbar\dot\Psi = H_A\Psi,
 \label{regSchrod}
 \end{equation}
where $H_A= -\frac{\hbar ^2}{2m^*}\nabla _A^*\nabla_A$ is determined from the vector potential of the magnetic field. Here, the geometric domain is the flat plane with the Cartesian coordinate system $(x,y)$. 
The purpose of this article is to examine the possibility that the planar system feeds back into the field in a certain particular way, as suggested by the Mesoscopic Mechanics (MeM). More precisely, the interaction is described by the following model. First, let $\Phi$ be a dimensionless complex-valued function of $(x,y)$, and  let it satisfy the following Nonlinear Schr\"odinger equation
\begin{equation}
\label{nlSch}
i\hbar\dot{\Phi }= H_A \Phi +\beta B^2/{\Phi ^*},
\end{equation}
We postulate that the magnetic field is determined by the modulus of $\Phi$ via the following relation:
\begin{equation}
\label{nlSch_0}
F_A=dA = \frac{B}{|\Phi|}dx\wedge dy.
\end{equation}
The two equations are considered jointly as a system. The vector potential is determined via (\ref{nlSch_0}) and re-enters (\ref{nlSch}) via the magnetic Hamiltonian. $B$ represents the magnetic field. Here, the constant $\beta$ is required in order to adjust the physical units. Since, up to a constant, $B^2$ is the energy-density, $\beta$ must incorporate a characteristic scale,  e.g. having to do with the material purity. The physical interpretation of $\Phi$ is beside the point as it is not used for itself. The role of $\Phi$ is to mediate between the electronic constituent of the system and the field, so as to establish the \emph{effective Hamiltonian }$H_A$. If the pair $(A, \Phi )$ satisfies the system (\ref{nlSch})--(\ref{nlSch_0}), then the corresponding $H_A$ is taken to be the effective Hamiltonian. I will comment on the source of this structure in the closing paragraphs of this article. 

Let us now shed some light on the nature of solutions of the system (\ref{nlSch})--(\ref{nlSch_0}). First, a standard calculation shows that the system is gauge-invariant. This is guarantied by the type of nonlinearity in Eq. (\ref{nlSch}) and the fact that Eq. (\ref{nlSch_0}) only involves the modulus of $\Phi$. Indeed, one checks directly that if the pair $(A,\Phi)$ satisfies the system, then so does the pair $(A+df, \Phi \exp{(-i\frac{e}{\hbar c}f)})$. In view of gauge-invariance, we can assume without loss of generality that the vector potential $A$ assumes the form 
\[
A=\delta\varphi = -\varphi _ydx+\varphi _xdy 
\]
for a certain real function $\varphi = \varphi (x,y)$. Furthermore, in view of  (\ref{nlSch_0}), $\varphi$ satisfies the Poisson equation
\begin{equation}
\label{lap}
\varphi _{xx}+\varphi _{yy} = B/|\Phi| .
\end{equation}
Subsequently, a calculation shows that $H_A$ assumes the following form:
\begin{equation}
2m^*H_A\Phi = -\hbar ^2(\partial ^2_x +\partial ^2_y)\Phi + 
2i\frac{e\hbar}{c}(-\varphi _y\Phi_x +\varphi _x\Phi _y)+\frac{e^2}{c^2}(\varphi _x^2+\varphi _y^2)\Phi .
\label{hamil}
\end{equation}
At this point, we make an Ansatz  
\[
\Phi(t,x,y) = e^{-iEt/\hbar}e^{ikx}\kappa (y),
\]
where $\kappa = \kappa (y)$ is a positive function, in fact $\kappa = |\Phi|$.
In such a case, (\ref{lap}) implies that we can further simplify the gauge representation of $A$ by asking that $\varphi =\varphi (y)$, so that in fact
\begin{equation}
\varphi _y = \int\limits_{-\infty}^y dy'B/\kappa (y').
\label{phint}
\end{equation}
Evaluating the Hamiltonian, we obtain
\begin{equation}
\begin{array}{rl}
2m^*e^{-ikx}e^{iEt/\hbar}H_A\Phi = & \hbar ^2k^2\kappa -\hbar ^2\kappa_{yy}  
+ 2\hbar k\frac{e}{c}\varphi _y\kappa+\frac{e^2}{c^2}\varphi _y^2\kappa \\
& = -\hbar ^2\kappa_{yy} +\left(\hbar k +\frac{e}{c}\varphi _y\right)^2\kappa .
\end{array}
\end{equation}
Since, at the same time,
$
i\hbar \dot\Phi = E\Phi,
$
the system (\ref{nlSch})-(\ref{nlSch_0}) will be satisfied if the real, positive function $\kappa=\kappa (y)$ satisfies the nonlinear integro-differential equation
 \begin{equation}
 -\kappa '' + \left(k +\frac{e}{\hbar c}\int\limits_{-\infty}^y dy'B/\kappa (y')\right)^2\kappa + \beta B^2\frac{2m^*}{\hbar ^2}\frac{1}{\kappa}= \frac{2m^*E}{\hbar ^2} \kappa .
\label {kap}
\end{equation}
Let us introduce a substitute variable
\[
s= k +\frac{e}{\hbar c}\varphi _y.
\]
Note first that in view of (\ref{lap}): $s'(y) = \frac{Be}{\hbar c}\kappa (y)^{-1}$,
so that \emph{$s'$ is proportional to the effective magnetic field} across any $y$-section of the plane. Recalculating Eq. (\ref{kap}) in the substitute variable, we obtain 
 \begin{equation}
 \frac{s'''}{s'} = 2\left(\frac{s''}{s'}\right)^2 -a s'^2-s^2+\mbox{E},
\label{s-equation}
\end{equation}
where 
\begin{equation}
a=\frac{2m^*c^2}{e^2}\beta [m^2] ,\quad \mbox{E}=\frac{2m^*}{\hbar ^2}E [m^{-2}].
\label{anE}
\end{equation}
We note that the unit of $s$ is $[m^{-1}]$. Also, our convention is such that the unit of $B$ is $[V/m]$ and the unit of $\beta$ is $[m^2C/V]$. Typical solutions of this nonlinear ODE are displayed in Fig. 1. Plots display $s'$, i.e. the derivative of the solution  of Eq. (\ref{s-equation}) (as well as the corresponding $s^2$---cf. next section for a comment) for four different choices of the re-scaled energy E and the parameter a. It assumes a bell-like, comb-like or double-delta-peak shape and is invariably concentrated on a narrow strip.

\section{Predictions---the qualitative picture}
\label{predic}

We will now examine the effective Hamiltonian $H_A=H_{A(s)}$. First, recalculating it with the use of variable $s$ we readily obtain
\begin{equation}
\frac{2m^*}{\hbar ^2}H_A\Psi = -(\partial ^2_x +\partial ^2_y)\Psi -2i(s-k)\Psi_x 
+(s-k)^2\Psi .
\label{effhamil}
\end{equation}
Also, the corresponding magnetic field is $\star F_A= \frac{\hbar c}{e}s'$.
Now, observe that $s$ and therefore $H_{A(s)}$ do not depend on $B$ explicitly. 
(Of course, the system remains sensitive to the total number of flux quanta carried by the field, although it would be clever to address this particular issue in a closed manifold setting rather than the \emph{unbounded} plane setting adopted here.) Let us now turn attention to the Schr\"odinger equation (\ref{regSchrod}) with the Hamiltonian given in (\ref{effhamil}). Consider candidate solutions in the form
\[
\Psi _n(t,x,y) = e^{-iE_nt/\hbar}e^{ikx}\chi _n (y),\quad \mbox{with the same $k$ as above.}
\]
In general, $\chi _n$ can be a complex valued function. Observe that $\Psi _n$ satisfies (\ref{regSchrod}) if and only if 
\begin{equation}
 -\chi _n '' +s^2\chi _n = \frac{2m^*}{\hbar ^2}E_n\chi _n .
\label{hamilout}
\end{equation}
One can only interpret $s^2=s^2(y)$ as the \emph{Hall potential} (when expressed in suitable units).  
Fig. 1 displays $s^2$ obtained by numerical simulation for a few distinct choices of the constants. In all cases, we observe a \emph{potential well} localized over a stripe-shaped region. The well is pinned to the area of concentration of the magnetic field. One is lead to expect that the spectrum $E_n$ consists of a discrete part corresponding to states trapped in the well as well as a continuous part corresponding to higher energy states moving to the right or to the left. This is in accord with the classical picture of charged particles getting trapped in the stripe of high magnetic field. Such localized states give rise to charge stripes stretching along the $x$-axis. What is extremely interesting is that the potential $s^2$ tends to display asymmetry. This is a very explicit manifestation of the Hall potential. Remarkably, the numerics suggest that the bias can be in either direction, depending on the regime, i.e. on the choice of $a$ and $\mbox{E}$.    

Let us find the heuristic formula for the Hall resistance. Let $L_x$ and $L_y$ denote the sample length and width. The superconducting current density is 
\[
J_n (x,y,t) =\frac{-e}{m^*}(L_xL_y)^{-1} \Re \left\{\Psi _n^*\left(i\hbar \nabla +\frac{e}{c}A \right)\Psi _n\right\}.
\]
Of course, the elementary charge may need to be adjusted, e.g. to $-2e$, etc., depending on the interpretation, but we will retain our original convention throughout this article. The total superconducting current is
\[
I_n  = \frac{e\hbar}{m^*}(L_xL_y)^{-1} \int\limits_{-\infty}^\infty s|\chi _n(y)|^2dy ,
\]
and it is entirely directed in the $x$-direction. Since the Hall voltage is 
$
V_H=\frac{\hbar ^2}{2m^*e}(s^2(\infty )-s^2(-\infty )), 
$
we find the Hall resistance 
\[
R_H= \frac{\hbar}{2e^2}L_xL_y\frac{s^2(\infty )-s^2(-\infty )}{\int s|\chi _n(y)|^2dy }.
\]
The exact values of Hall resistance in this model can only be established by careful numerical and theoretical analysis of equation (\ref{s-equation}) to be undertaken in the future.  Let us note that if the geometric space can be selected as a compact manifold, then topological quantization of the magnetic field comes into play. If such a manifold is cut along a ``junction", then the notion of Hall potential makes sense and can be discussed in a similar way as above. In such a scenario, quantization of the Hall resistance is expected.   

Various anomalies regarding the Hall effect, including negative Hall resistance, have been experimentally verified in all sorts of high-temperature superconductors, cf. e.g. \cite{hopf}, \cite{jin}, \cite{naro}. 

\section{The conceptual origin of the model}

The Hamiltonian corresponding to the Schr\"dinger equation (\ref{nlSch}) is:
\begin{equation}
\tilde\Xi (\Phi ) = \int \limits_{2D  plane}\left(\Phi ^* H_A \Phi  + \beta B^2\log |\Phi |^2\right).
\label{KsiP}
\end{equation}
Functionals of this type are at the foundation of the Mesoscopic Mechanics (MeM, \cite{sowa5}),  and the Nonlinear Maxwell Theory,   \cite{sowa4}. I have already shown that the MeM fully conforms with Quantum Mechanics, cf. \cite{sowa6}. However, the MeM also predicts that a novel type of interaction between an ensemble of correlated electrons and the ambient magnetic field may come into play in two-dimensional systems.  In the present article I have demonstrated via a simplified model how this type of interaction affects the electronic properties. 

The 2D system of equations presented here admits a consistent extension to $3+1$ dimensions, which is given by the nonlinear Maxwell system:
\begin{equation}
 dF_{A}=0, \quad
 d^* (\lambda\left(|\Phi |\right)F_{A})=0
\label{syst1}
\end{equation}
\begin{equation}
-i\hbar\dot{\Phi }= H_A \Phi +\beta |F_{A}|^{2}\Phi .
\label{syst2}
\end{equation}
The function $\lambda$ is either defined as $\lambda (x)=x$, or 
$
\lambda (x) = (c-\ln{|x|})^{-1}
$ with a suitable constant $c$. In both cases the system has the same two-dimensional reduction, i.e. (\ref{nlSch})-(\ref{nlSch_0}). However, only in the second case the system can be tied to a variational principle. I have examined equations of this type in previous articles, cf. \cite{sowa4}, where the dispersion relation in the Schr\"odinger part of the equations is set to be the classical dispersion relation of the wave equation. Also, articles \cite{sowa1}, and \cite{sowa3} are particularly closely related to the present work. In those papers the Hamiltonian is simplified to the Laplacian and the order parameter is traded for a time-independent real function (in our present notation, it would correspond to the modulus $|\Phi |$).

\newpage
\noindent
\textbf{Fig. 1.} Plots of $s'$, i.e. the derivative of the solution  of Eq. (\ref{s-equation}) and the corresponding transversal potential $s^2$  for four different choices of the re-scaled energy E and the parameter a. The function $s'$ is proportional to the density of the effective magnetic field. It assumes a bell-like, comb-like or double-delta-peak shape and is invariably concentrated on a narrow strip. The Hall potential given by $s^2$ can be biased in the regular or anomalous direction. In all cases, there is a prominent potential well aligned with the stripe of high magnetic field. This is in accord with the classical picture of charged particles getting trapped in the high magnetic field area. [Numerical simulations have been generated with the Scilab software package of INRIA]

\vbox spread 1.5in{}
\includegraphics{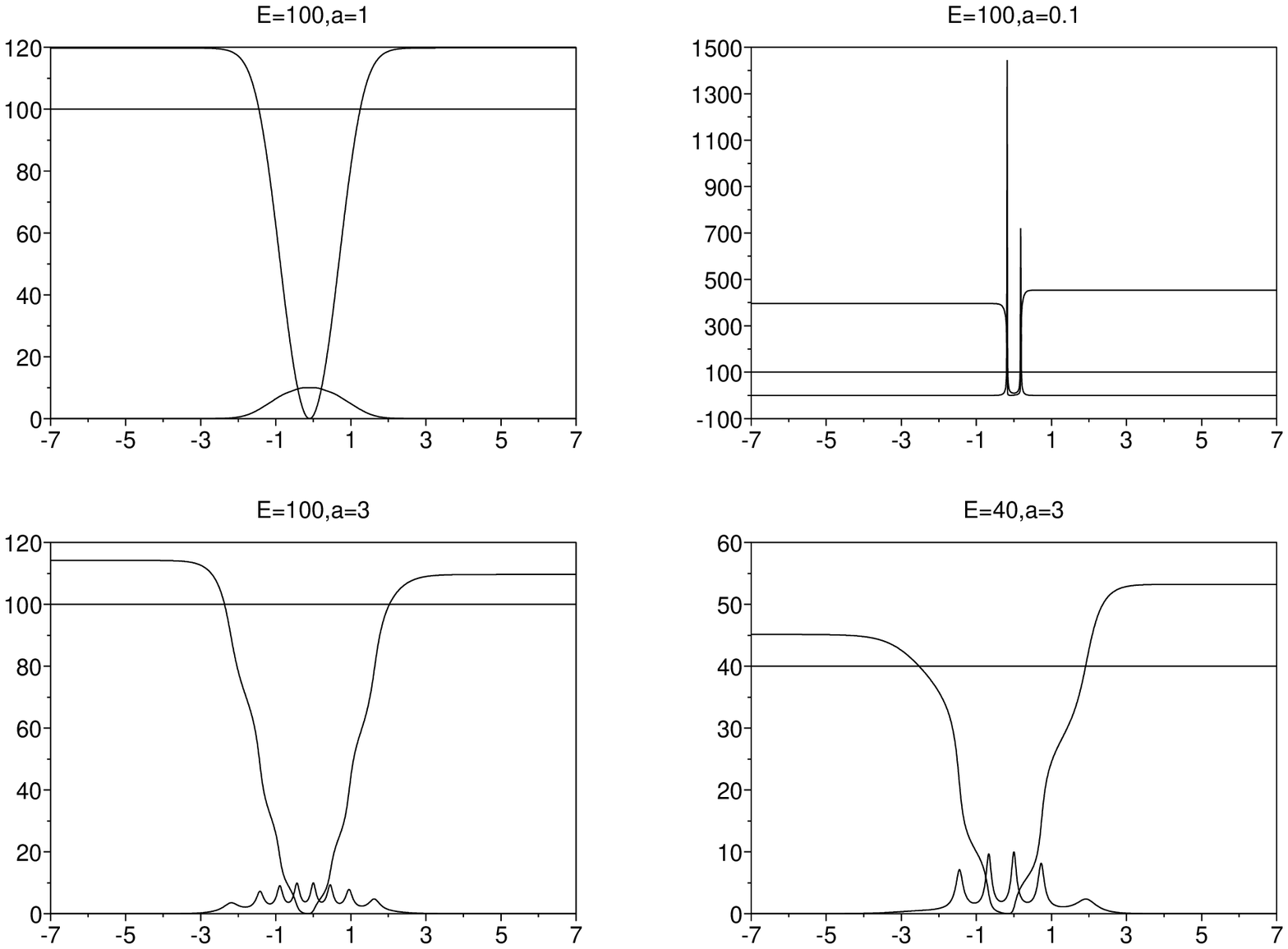}

\end{document}